%% file: main.tex
\pdfoutput=1
\documentclass[runningheads]{llncs}
\usepackage[T1]{fontenc}
\usepackage{orcidlink}
\usepackage{graphicx}
\usepackage[inkscapelatex=false]{svg}
\usepackage{setspace}
\usepackage{times}
\usepackage{amssymb}
\usepackage{amsmath}
\usepackage{graphicx}
\usepackage{float}      
\usepackage{caption}
\usepackage{booktabs}
\usepackage{subcaption}
\usepackage{adjustbox}
\usepackage{algorithm}
\usepackage{algpseudocode}
\usepackage{placeins}
\usepackage{tcolorbox}
\usepackage{hyperref}
\usepackage{xurl}
\usepackage[numbers,sort&compress]{natbib}
\tcbuselibrary{skins,breakable}

\begin{document}
\title{Bitcoin After Block Rewards}
%
%
\author{Junhyuk Lee\orcidlink{0009-0002-9490-2808}}
\authorrunning{J. Lee}
%
\institute{Texas A\&M University, College Station, TX, USA \\ \email{xodn348@tamu.edu}}

\maketitle              
\begin{abstract}
Bitcoin’s block reward is scheduled to decline to zero, raising concerns about
whether the network can remain secure once miners rely solely on transaction fees.
This paper seeks to identify the conditions under which large-scale and persistent
deviation from honest mining can arise.

We analyze and compare the payoffs of honest and deviating miners in a sequential
decision model, and identify a deviation threshold $G_t$ at which honest mining
ceases to be privately optimal.
Around the 2024 Bitcoin halving, we show that current mining
behavior does not exhibit large-scale or structural deviation.
However, when the block reward is removed, the $G_t$ criterion implies that
deviation can arise even with a very small fraction of transaction fees.

Finally, we evaluate three protocol-level mechanisms: Base Fee, Fee Floor, and an
adaptive maximum block size rule, and show that their combination raises the
deviation threshold and mitigates incentive breakdown in a fee-only regime.
These results provide a practical benchmark for assessing Bitcoin’s security as
block rewards disappear.

\end{abstract}

\section{Introduction}

\subsection{Motivation}

Bitcoin was introduced by Satoshi Nakamoto as a decentralized alternative to traditional financial infrastructures~\cite{Bitcoin}, motivated in part by concerns regarding centralized control and discretionary intervention during the 2008 financial crisis. The system operates on a distributed network of nodes and maintains its security through the collective participation of miners, who contribute computational resources known as hashing power to validate transactions and prevent attacks such as double spending or unauthorized ledger modifications.

The security of the Bitcoin network critically depends on these mining incentives. However, Bitcoin’s mining protocol dictates that block rewards halve approximately every four years in order to enforce long-term scarcity. This design implies a corresponding decline in miner revenue unless compensated by increases in external factors such as transaction fees or market demand. As block rewards diminish, the risk arises that miners may withdraw hashing power, thereby reducing the cost of attacks and threatening the network’s long-term security.

In the post-reward era, miner compensation must rely primarily on transaction fees and emerging sources of value such as Maximum Extractable Value (MEV). These components increase uncertainty in miner revenue and create additional strategic considerations for miners. An unstable reward structure can create misaligned incentives, encouraging miners to deviate from honest mining to maximize short-term gains. Such deviations may compromise consensus threshold and contribute to network security instability.

These concerns motivate this research, which investigates how Bitcoin’s protocol and incentive mechanisms can evolve to sustain network security once block rewards are no longer available. Understanding and formalizing miner behavior in the post-reward era is essential for evaluating the long-term viability of Bitcoin as the first digital currency and for preserving the financial
autonomy that underlies its original design.

\subsection{Problem Statement}
The central question of this thesis is whether the Bitcoin network can remain secure when miner
revenue is derived solely from transaction fees. In particular, we examine
whether miners have sufficient incentives to continue honest mining in a fee-only environment and
whether such conditions satisfy the Byzantine Fault Tolerance (BFT) stability threshold required for network security.

\subsection{Objectives}
This thesis studies whether Bitcoin can remain secure as block reward declines and miner
revenue becomes increasingly dependent on transaction fees and potential private gains.
The core objectives are as follows:

\begin{enumerate}
    \item Formulate a per-block miner decision model that characterizes the choice between
    honest mining and deviation through the deviation threshold condition
    $G_t \ge \phi(w) \cdot X_t$.

    \item Empirically analyze the 2024 halving window (block heights 790{,}000--890{,}000) to
    examine how reductions in observable rewards $X_t$ and changes in fee-market conditions
    affect miner profitability and deviation-consistent behavior, using block audit scores as a proxy.

    \item Evaluate miner profit–stabilization mechanisms, including a base fee, a fee floor, and an adaptive maximum block size rule under a hypothetical fee-only regime, and identify policy combinations that keep the deviation rate below the BFT-style threshold.
\end{enumerate}

\subsection{Contributions}
This thesis makes the following contributions to understanding Bitcoin miner behavior
as block reward declines.

\begin{enumerate}
    \item It shows that rational miner deviation can be characterized by a simple per-block
    threshold linking additional private gains to observable miner rewards, clarifying how
    reductions in $X_t$ mechanically relax deviation constraints in a fee-dominated environment.

    \item It reveals that although observable miner rewards and on-chain demand declined
    substantially, deviation-consistent behavior did not increase persistently while block
    rewards remained positive, indicating the stabilizing role of the remaining block reward.

    \item It demonstrates that in a hypothetical fee-only regime, deviation becomes widespread
    even for small private gains, and proposes reward stabilization mechanisms,
    including a base fee, a fee floor, and an adaptive maximum block size, that mitigate massive
    deviation and help maintain BFT-style security thresholds.    
\end{enumerate}

\section{Related Work}

\subsection{Miner Incentives and Deviation}

Bitcoin’s security relies on miners being economically incentivized to extend the longest chain,
yet prior work shows that this incentive structure becomes fragile as block rewards decline and
mining revenue increasingly depends on transaction fees. Carlsten et al.~\cite{instability} analyze
a fee-dominated regime and show that high variance in transaction fees gives rise to
incentive-compatible deviations such as fee-sniping and delayed mining, highlighting the
stabilizing role of block subsidies in discouraging deviation. Complementing this view, Eyal and
Sirer~\cite{selfish_mining} demonstrate that strategic deviation can be profitable even without a majority
of hashing power when network propagation is imperfect, as withholding and selective block release
allow deviating miners to earn rewards exceeding their proportional share. Together, these results
establish that reduced block rewards and propagation delays jointly weaken incentives for honest
mining, motivating a formal analysis of deviation conditions.

\subsection{MEV and Block Propagation Delay on Mining Incentives}

Deviation gains arise when miners do not receive blocks and transactions at the same time.
Daian et al.~\cite{flashboys2020} show that small timing advantages in information propagation can
be directly monetized through MEV, while Decker and Wattenhofer~\cite{decker2013information}
empirically show that block propagation delay is common and is the primary cause of blockchain
forks and orphaned blocks. As propagation delay increases, honest miners waste more work on blocks
that do not enter the main chain, while miners who can exploit or tolerate such delays are less
affected, widening the payoff gap between honest and deviating behavior.

\subsection{Consensus Security}

While the above works focus on miner incentives, consensus-security analyses formalize the
conditions under which the Bitcoin protocol remains safe and live. Garay, Kiayias, and Leonardos
develop the Bitcoin Backbone model~\cite{bitcoin_backbone,DBLP:conf/crypto/GarayKL17}, which proves that the protocol achieves
robust chain growth and chain quality as long as the adversarial hashing power remains below a given
fraction of the total. Their framework highlights the importance of maintaining an honest majority
and provides a probabilistic justification for using a Byzantine Fault Tolerance style threshold
when evaluating network stability.

\subsection{Relevant Bitcoin Improvement Proposals (BIPs)}
As summarized in Table 1, several Bitcoin Improvement Proposals (BIPs) have explored changes to Bitcoin’s block-size
limit, primarily to accommodate growing transaction demand. Early proposals such as
BIP-100, BIP-101, and BIP-102 advocated increasing the maximum block size through fixed
expansions or scheduled growth, focusing mainly on throughput and congestion relief
\cite{BIP100,BIP101,BIP102}. 

Later proposals introduced adaptive mechanisms that adjust the block-size limit based on
network conditions. BIP~103 and BIP~104 propose algorithmic rules tied to bandwidth growth
or block difficulty, while BIP~107 adjusts the limit according to recent block utilization
\cite{BIP103,BIP104,BIP107}. These approaches aim to improve scalability but do not
explicitly consider how block-size dynamics affect miner deviation incentives.
\vspace{-1.5em}
\begin{table}[H]
\centering
\caption{Summary of Block-Size Related Bitcoin Improvement Proposals (BIPs)}
\begin{adjustbox}{width=\textwidth}
\begin{tabular}{|c|l|l|p{6.5cm}|}
\hline
\textbf{BIP} & \textbf{Author(s)} & \textbf{Category / Mechanism} & \textbf{Key Idea} \\ \hline

100 & Garzik et al. &
Miner voting / Dynamic sizing \cite{BIP100} &
Miners vote to adjust the maximum block size periodically. \\ \hline

101 & Andresen &
Fixed schedule  \cite{BIP101} &
Increase block size to 8 MB and double every two years to follow projected demand. \\ \hline

102 & Garzik &
Fixed increase  \cite{BIP102} &
One-time increase to a 2 MB block size. \\ \hline

103 & Wuille &
Algorithmic / Bandwidth-based \cite{BIP103} &
Adjust block size according to long-term bandwidth growth (17.7\% per year). \\ \hline

104 & Khan &
Algorithmic / Difficulty-like rule \cite{BIP104} &
Block-size cap adjusts similarly to mining difficulty using 75\% target utilization. \\ \hline

105 & BtcDrak &
Algorithmic / Retargeting algorithm \cite{BIP105} &
Consensus-driven dynamic block-size retargeting mechanism with ±10\% adjustment window. \\ \hline

106 & Chakraborty &
Algorithmic / Automatic control \cite{BIP106} &
Automatically controlled block-size limit responding to previous block size or transaction-fee conditions. \\ \hline

107 & Sanchez &
Algorithmic / Utilization-based \cite{BIP107} &
Block-size limit is adjusted based on recent block utilization trends. \\ \hline

109 & Andresen &
Fixed + policy constraints \cite{BIP109} &
Increase block size to 2 MB and introduce signature-operation limits for validation stability. \\ \hline

\end{tabular}
\end{adjustbox}
\label{tab:bip-summary}
\end{table}

\section{Model Assumptions}

\subsection{Miner Computation and Pool Participation}
We model each miner as an independent computational agent executing an
\textbf{interactive Turing machine (ITM)}.
This abstraction captures the fact that mining is an ongoing, interactive
process in which miners continuously receive information from the network,
such as new blocks and transactions, and adapt their behavior accordingly.

Miners may choose to participate in mining pools, but pool participation is
treated as a voluntary decision made solely to maximize individual profit.
Regardless of whether a miner operates solo or within a pool, rewards are
assumed to be distributed proportionally to the miner’s contributed hashing
power.
As a result, joining a pool does not alter the miner’s underlying strategic
objective, but only affects the aggregation and sharing of rewards.

Accordingly, we analyze strategic behavior at the level of individual miners
and do not model pools as independent strategic agents.
Collective pool behavior is treated as the aggregate outcome of individual
miners’ decisions, and group-level strategy is therefore outside the scope
of this model.

\subsection{Rational Miners as Long-Lived Strategic Agents}
We model miners as rational, long-lived strategic agents.
Current mining decisions can influence future decision environments, and the
economic consequences of mining behavior such as profit realization,
operational costs, and risk exposures are determined along a miner’s own
timeline rather than at an isolated block level.
As a result, miners do not make decisions on a purely block-by-block basis,
but instead account for how present actions affect future outcomes.

For this reason, miner behavior is modeled using a \textbf{Markov Decision Process
(MDP)}, which is the standard framework for sequential decision-making in
stochastic and evolving environments~\cite{suttonbarto2018}.
The MDP formulation allows mining strategies to be evaluated not only by their
immediate payoff, but also by their impact on future states and expected
profits, which is consistent with long-lived strategic behavior in blockchain
systems.

The objective of miner $i$ is defined as
\begin{equation}
V_i = \mathbb{E}\!\left[ \sum_{t=0}^{\infty} \gamma^t \, \Pi_i(S_t, a_i(t)) \right],
\end{equation}
where $t$ indexes block discovery window following the resolution of the fork
competition for block $t-1$, $\Pi_i(S_t, a_i(t))$ denotes the per-block profit,
and $\gamma \in (0,1)$ is a discount factor.

At block $t$, the network state observed by miner $i$ is defined as
\begin{equation}
S_t = \bigl( F_t,\; M_t,\; h_i,\; \delta_t \bigr),
\end{equation}
where $F_t$ denotes the transaction fee revenue observed at $t$, $M_t$
denotes the realized miner-extractable value (MEV), $h_i$ is miner $i$'s share
of total network hashing power, and $\delta_t$ represents the effective network
propagation delay determined by block size and mining strategy.
The action of miner $i$ at block $t$ is given by
\begin{equation}
a_i(t) \in \{\text{hon},\; \text{dev}\},
\end{equation}
corresponding to honest mining and deviating mining strategies, respectively.

Miners evaluate mining strategies over long-term investment horizons,
while decisions are made on a per-block basis.
The MDP formulation is therefore not used to assume myopic block-by-block
decision-making, but to justify it: given the Markov structure of the state
and payoff, the optimal policy admits a per-block decision rule that maximizes
expected long-term discounted payoffs for a long-lived strategic agent.
\label{sec:MDP}

\subsection{Honest Mining and Deviating Mining}

At each block, miners choose between honest and deviating mining strategies. 
We assume that deviation does not alter a miner’s computational capability; 
that is, the miner’s hash rate remains unchanged across strategies, 
and only the propagation behavior differs.

\textbf{Honest mining} refers to protocol-conformant behavior in which miners
follow the Bitcoin protocol as specified.
Honest miners extend the longest known valid chain, construct blocks using
transactions from the public mempool, and immediately broadcast newly
discovered blocks to the network.
Operationally, this corresponds to repeatedly attempting to solve the PoW
puzzle for the current chain tip and, upon success, diffusing the resulting
block without delay, as specified by the standard Bitcoin protocol and its
formalizations~\cite{Bitcoin}.

\textbf{Deviating mining} refers to any strategy that departs from this
protocol-conformant behavior in order to increase private profit.
Examples include block withholding, delayed block publication, and selective
transaction inclusion to capture higher fees or other private gains.
Such strategies intentionally deviate from immediate block propagation or
standard block construction rules to improve expected payoff~\cite{selfish_mining}.
\label{sec:honest-deviating}

\subsection{Partially Synchronous Network Model}
We assume a partially synchronous network model in the sense of
Dwork, Lynch, and Stockmeyer~\cite{dls88},
where there exists an unknown but finite upper bound on message
delivery delay that eventually holds.
This model captures realistic network behavior that lies strictly
between fully synchronous and fully asynchronous communication.

We model block propagation using an effective delay function
$\delta(B)$, which depends on block size and network conditions.
We assume that
\begin{equation}
\delta(B) \le \Delta \qquad \text{for all block sizes } B,
\end{equation}
where $\Delta$ denotes an unknown but finite network delay bound.
This assumption guarantees eventual block propagation while allowing
temporary forks and orphaned blocks induced by propagation latency.

\subsection{Block Propagation Delay}

We model block propagation delay as an increasing function of block size.
Empirical measurements show that block propagation delay grows approximately
linearly with block size due to bandwidth and validation constraints in the
peer-to-peer network~\cite{croman2016scaling}.

Let $\delta(B)$ denote the propagation delay of a block with size $B$.
The delay is modeled as
\begin{equation}
\delta^{\mathrm{hon}}(B) = \delta_0 + \kappa \cdot B,
\end{equation}
where $\delta_0$ is the baseline block propagation delay and $\kappa$ captures the
incremental delay per unit of block size.
Throughout the analysis, block size is measured in virtual bytes (vB),
reflecting the SegWit-adjusted block weight.

For deviating mining strategies, we allow miners to intentionally delay block
publication.
Accordingly, the propagation delay under deviation is given by
\begin{equation}
\delta^{\mathrm{dev}}(B) = \delta^{\mathrm{hon}}(B) + w,
\end{equation}
where $w \ge 0$ represents an intentional withholding period chosen by the
miner.
This additional delay increases the probability of orphaning and directly
affects the expected payoff of deviating strategies. Additional examination of the withholding time $w$ is provided in the Appendix.
\label{sec:block-propagation-delay}

\subsection{Successful Mining and Propagation}
We define the successful mining probability of miner $i$ as the
probability that a block discovered by the miner is not orphaned and is
successfully incorporated into the main chain.
This probability accounts for both block discovery and successful network
propagation.

Let $h_i$ denote miner $i$'s share of total network hashing power, and let
$\rho$ denote the orphan rate.
The successful mining probability is then given by
\begin{equation}
p_i(t) = h_i (1 - \rho_t)
\end{equation}
The block discovery rate $h_i$ remains unchanged across strategies, 
as we assume that miners have identical computational capabilities 
under both honest and deviating mining. Also, the orphan rate $\rho$ captures the probability that a competing block is
discovered during the propagation window of a newly mined block.
Consistent with prior analytical and empirical studies of Bitcoin mining,
block discovery is modeled as a \textbf{Poisson process} with arrival rate
$\lambda$~\cite{decker2013information}.
If $\delta$ denotes the block propagation delay, the probability that at
least one competing block is discovered during this interval is
\begin{equation}
\rho_t = 1 - \exp(-\lambda \delta_t)
\end{equation}
Accordingly, a block is successfully mined only when no competing block is
found during its propagation period. 
\label{sec:successful-mining-propagation}

\subsection{Protocol Summary}
\label{sec:protocol-summary-itm}

\begin{tcolorbox}[
  colback=white,
  colframe=black,
  fonttitle=\bfseries,
  title={Mining for block $t$},
  breakable
]

\textbf{Parties.}
Individual miners $\{1,\dots,n\}$, each modeled as an interactive Turing machine (ITM).

\medskip
\textbf{Input (to miner $i$).}
Network state $S_t = (F_t, M_t, h_i, \delta_t)$ and miner cost $C_i$.

\medskip
\textbf{Execution.}
For each block $t$, miner $i$ executes the following steps:
\begin{enumerate}
  \item \textbf{Observe.}
  Observe the network state $S_t$, which summarizes the relevant information
  associated with the block discovery event at block $t$.

  \item \textbf{Decide.}
  Choose a mining strategy for block $t$ based on the observed state $S_t$,
  \[
    a_i(t) \in \{\mathrm{hon}, \mathrm{dev}\},
  \]
  corresponding to honest mining or deviating mining behavior.

  \item \textbf{Execute.}
  Execute the chosen mining strategy.
  If $a_i(t)=\mathrm{hon}$, follow honest mining behavior.
  If $a_i(t)=\mathrm{dev}$, follow deviating behavior (e.g., withholding or
  selective transaction inclusion) as described in Section~\ref{sec:honest-deviating}.

  \item \textbf{Propagate.}
  Upon block discovery, broadcast the block.
  The block is subject to an effective propagation delay according to the delay
  model in Section~\ref{sec:block-propagation-delay}, which probabilistically determines the block outcome
  (adopted or orphaned) as described in Section~\ref{sec:successful-mining-propagation}.
\end{enumerate}

\medskip
\textbf{Output.}
A realized block outcome (adopted or orphaned) and the corresponding realized
payoff $\Pi_i(S_t, a_i(t))$ as defined in Section~\ref{sec:MDP}.

\end{tcolorbox}

\section{Problem Formulation}

In this chapter, we formalize Bitcoin miners’ profit incentives by defining the total profit function $V_i$
and the per-block payoff function $\Pi_i$.
We then characterize the additional deviation gain $G_t$ required for deviation to become
economically rational and derive the corresponding threshold condition under which deviation occurs.
By controlling the magnitude of $G_t$, the fraction of miners for whom deviation is profitable can be reduced,
thereby maintaining the network’s fault-tolerance threshold of $1/2$ as established in the Bitcoin
Backbone protocol~\cite{bitcoin_backbone,DBLP:conf/crypto/GarayKL17}.

\subsection{Miners' Profit}
We model miners as rational agents who maximize long-run discounted profit~\cite{suttonbarto2018}. 
The total profit of miner $i$ is defined as
\begin{equation}
V_i \;=\; \mathbb{E}\!\left[\sum_{t=0}^{\infty} \gamma^{t} \, \Pi_i(S_t, a_i)\right],
\label{MDP}
\end{equation}
where $\Pi_i(S_t, a_i)$ denotes the per-block payoff at network state $S_t$ under action $a_i$, 
and $\gamma \in (0,1)$ is a discount factor. 
Throughout this work, we set the discount factor per block interval (roughly 10 minutes) to
\(\gamma = 0.99^{1/144} \approx 0.99993\),
so that the implied daily discount factor is \(0.99\).

The per-block payoff function is defined as
\begin{equation}
\Pi_i(S_t, a_i) \;=\; p_i \cdot X_t - C_i,
\end{equation}
where $p_i$ is the successful mining probability of miner $i$, 
$X_t$ is the total reward available in block $t$, 
and $C_i$ represents the miner’s operational cost. 
The total reward $X_t$ is given by
\begin{equation}
X_t = R_t + F_t + M_t,
\end{equation}
where $R_t$ is the block reward, $F_t$ denotes transaction fees, and $M_t$ captures miner extractable value (MEV).
In the post-reward regime considered in this chapter, we assume $R_t = 0$, so that
$X_t = F_t + M_t$.
The cost term $C_i$ represents the mining cost incurred in producing a block, including electricity expenses and hardware depreciation.

Although miners are long-lived strategic agents with $\gamma = 0.99993$,
both honest and deviating strategies share the same long-run value function,
and the state is Markov. The optimal policy therefore reduces to a per-block
decision rule, without assuming myopic behavior.
At each block $t$, a miner chooses between honest mining and deviation by comparing the corresponding per-block profits. Let $\Pi_i^{\text{hon}}(t)$ and $\Pi_i^{\text{dev}}(t)$ denote the per-block payoffs under honest and deviating behavior, respectively.
These are given by
\begin{align}
\Pi_i^{\text{hon}}(t) &= p_i^{\text{hon}} \cdot X_t - C_i, \\
\Pi_i^{\text{dev}}(t) &= p_i^{\text{dev}} \cdot (X_t + G_t) - C_i,
\end{align}
where $G_t$ represents the additional gain obtainable only through deviation, such as withholding or opportunistic forking.

\begin{algorithm}[h]
\caption{Miner Decision at Block $t$}
\label{alg:block-decision}
\begin{algorithmic}[1]
\For{each block $t$}
    \For{each miner $i$}
        \If{$\Pi_i^{\mathrm{dev}}(t) \geq \Pi_i^{\mathrm{hon}}(t)$}
            \State $a_i(t) \gets \mathrm{deviation~mining}$ \Comment{forking or withholding}
        \Else
            \State $a_i(t) \gets \mathrm{honest~mining}$
        \EndIf
    \EndFor
\EndFor
\end{algorithmic}
\end{algorithm}

\subsection{Additional Deviation Gain $G_t$}
At each block $t$, miner $i$ compares the per-block payoffs of honest mining and deviation. This per-round comparison is justified because the underlying value function
$V_i(\cdot)$ is identical for honest and deviating strategies.
The strategies differ only in the realized per-block payoff $\Pi_i$, which allows
the miner to directly compare $\Pi_i^{\mathrm{hon}}(t)$ and
$\Pi_i^{\mathrm{dev}}(t)$ at each block.

The miner $i$ chooses deviation whenever
\begin{align}
\Pi_i^{\mathrm{dev}}(t) \geq \Pi_i^{\mathrm{hon}}(t).
\end{align}
Substituting the payoff definitions yields
\begin{align}
p_i^{\mathrm{dev}}(X_t + G_t) 
&\geq p_i^{\mathrm{hon}} X_t, \\
p_i^{\mathrm{dev}}X_t + p_i^{\mathrm{dev}}G_t 
&\geq p_i^{\mathrm{hon}}X_t, \\
p_i^{\mathrm{dev}}G_t 
&\geq (p_i^{\mathrm{hon}} - p_i^{\mathrm{dev}})\,X_t, \\
G_t 
&\geq \frac{p_i^{\mathrm{hon}} - p_i^{\mathrm{dev}}}{p_i^{\mathrm{dev}}}\, X_t.
\end{align}
We define the multiplicative term on the right-hand side as
\begin{align}
\phi(w)
&\triangleq \frac{p_i^{\mathrm{hon}} - p_i^{\mathrm{dev}}}{p_i^{\mathrm{dev}}},
\end{align}
which represents the relative success-probability penalty induced by deviation. 
The term $\phi(w)$ arises from the withholding time $w$, and increases as $w$ grows. 
A larger withholding time reduces the miner’s effective success probability 
while increasing the orphan rate, thereby requiring a higher deviation gain $G_t$ 
to make deviation profitable. Accordingly, we refer to $\phi(w)$ as the \textit{withholding penalty factor}.
The $\phi(w)$ will be dealt with in Further Discussion chapter later. Using this definition, the incentive condition can be written compactly as
\begin{align}
G_t 
&\geq \phi(w) \cdot X_t.
\end{align}

As shown above, the deviation threshold for $G_t$ is proportional to both the withholding penalty factor
$\phi(w)$ and the total available reward $X_t$.
Deviation can therefore be discouraged either by reducing the deviation-only gain $G_t$
in the left-hand side or by increasing threshold through a larger $\phi(w)$ or $X_t$ in the right-hand side.
However, $\phi(w)$ is primarily determined by miner-specific behavior and network
conditions, such as withholding time and propagation delay, and is not directly controllable
by the protocol.
For this reason, we treat $\phi(w)$ as an external cause and focus our analysis
on mechanisms that reduce $G_t$ and raise $X_t$.

\section{Empirical Data Analysis}
\subsection{Overview}
The goal of this section is to examine how changes in observable rewards
$X_t = R_t + F_t + M_t$ affect the deviation threshold
$G_t \ge \phi(w) \cdot X_t$, and to inspect whether a reduction in
$X_t$ influences deviation mining and induces significant strategic changes
by miners, including the halving event. 

\subsection{Empirical Setup and Measurement}

    \paragraph{\textbf{Data source.}}
    Bitcoin on-chain data are obtained from blockchain.com, and block-level audit
    metrics are sourced from mempool.space.
    
    \paragraph{\textbf{Data type.}}
    From blockchain.com, we extract block-level statistics including total fees
    (in satoshis), total transaction size (vBytes), block weight, and mining pool
    identifiers. Block audit scores at each block height are obtained from
    mempool.space.

    \paragraph{\textbf{Data calibration.}}
    Throughout the observation window, median values are used for $X_t$, $F_t$,
    fee rate (sat/vB), and block utilization ratios to mitigate the impact of
    outliers and to capture underlying trends in each metric.
    
    \paragraph{\textbf{Experimental window.}}
    The 100{,}001-block dataset spans block heights 790{,}000 to 890{,}000,
    corresponding to a roughly $\pm 50{,}000$ block window around the 2024 halving at
    height 840{,}000. This period covers May 16, 2023 to March 29, 2025, and
    includes both pre-halving and post-halving regimes.
        
    \paragraph{\textbf{Revenue and cost.}}
    All revenue and cost quantities are computed at a daily frequency.
    For each block, the miner reward consists of the block subsidy $R_b$,
    transaction fees $F_b$, and MEV $M_b$.
    Daily revenue for mining pool $i$ is obtained by aggregating block-level rewards
    and multiplying by the BTC--USD price on day $d$.
    
    Mining costs are estimated using the annualised electricity consumption estimate
    $\mathrm{GUESS}(d)$ (TWh/year) reported by the Cambridge Bitcoin Electricity
    Consumption Index (CBECI).
    Following the CBECI methodology, which models electricity expenditure for
    industrial-scale mining operations using a constant global average electricity
    price~\cite{cbe201:}, we assume an electricity price of
    $p_e = 0.05$ USD/kWh.
    We convert annual consumption to a daily network-wide electricity cost by
    dividing by 365.25 and multiplying by $p_e$.
    The resulting daily network cost is then allocated to each mining pool in
    proportion to the number of blocks it mines on day $d$.
    Daily profits are finally aggregated into monthly time series.
    \newcommand{\eqdesc}[1]{\makebox[0.4\linewidth][l]{#1}}
    \begin{align}
    \eqdesc{Daily BTC reward of pool $i$ :}
    & X_i(d) = \sum_{b \in \mathcal{B}_{i,d}} \bigl(R_b + F_b + M_b\bigr), \\
    \eqdesc{Daily revenue in USD :}
    & \mathrm{Rev}_i(d) = P_{\mathrm{BTC}}(d)\, X_i(d), \\
    \eqdesc{Daily network electricity cost :}
    & C_{\mathrm{net}}(d) = \frac{\mathrm{GUESS}(d)}{365.25}\, p_e, \\
    \eqdesc{Electricity cost per block :}
    & c(d) = \frac{C_{\mathrm{net}}(d)}{N(d)}, \\
    \eqdesc{Daily electricity cost of pool $i$ :}
    & C_i(d) = c(d)\, n_i(d), \\
    \eqdesc{Daily profit of pool $i$ :}
    & \Pi_i(d) = \mathrm{Rev}_i(d) - C_i(d).
    \end{align}

    \paragraph{\textbf{Assumption in Max Extractable Value (MEV).}}
    Because there is no widely accepted empirical measurement of MEV in the Bitcoin
    network, we approximate $M_t$ by conservatively scaling empirical MEV measurements from Ethereum to reflect Bitcoin’s more limited transaction structure and lower complexity~\cite{blockscholes2024mev,flashbots2022quantifying}.
    We model $M_t$ using a zero-inflated lognormal distribution and generate 100{,}001 synthetic MEV samples, assigning one sampled value to each block in our analysis. MEV is treated as an exogenous component and is added to transaction fees when computing miner payoffs. This simplified treatment captures occasional private gains without making MEV a primary driver of the results.
        
    \paragraph{\textbf{Deviation indicator.}}
    Because miner deviation strategies are not directly observable from public
    blockchain data, we use the block audit score as a proxy for
    deviation-consistent behavior. The audit score measures the ratio of actual
    miner revenue to the expected revenue under a public-mempool
    fee-maximizing template. Lower audit scores indicate departures from
    public-mempool fee maximization, which may arise from the presence of private
    transaction fees or other forms of private gains $G_t$. Simply put, a low
    audit score implies that the miner did not follow the public mempool’s
    expected fee-maximizing protocol.
    \begin{equation}
    \mathrm{Audit~Score}
    =
    \frac{\text{actual miner revenue in the block}}
         {\text{expected revenue under the public-mempool fee-maximizing template}}.
    \end{equation}

\subsection{Analysis Result}
\subsubsection{Decline of $R_t$ and $F_t$ in $X_t$}

\begin{figure}[p]  
    \centering
    \includegraphics[width=1\linewidth]{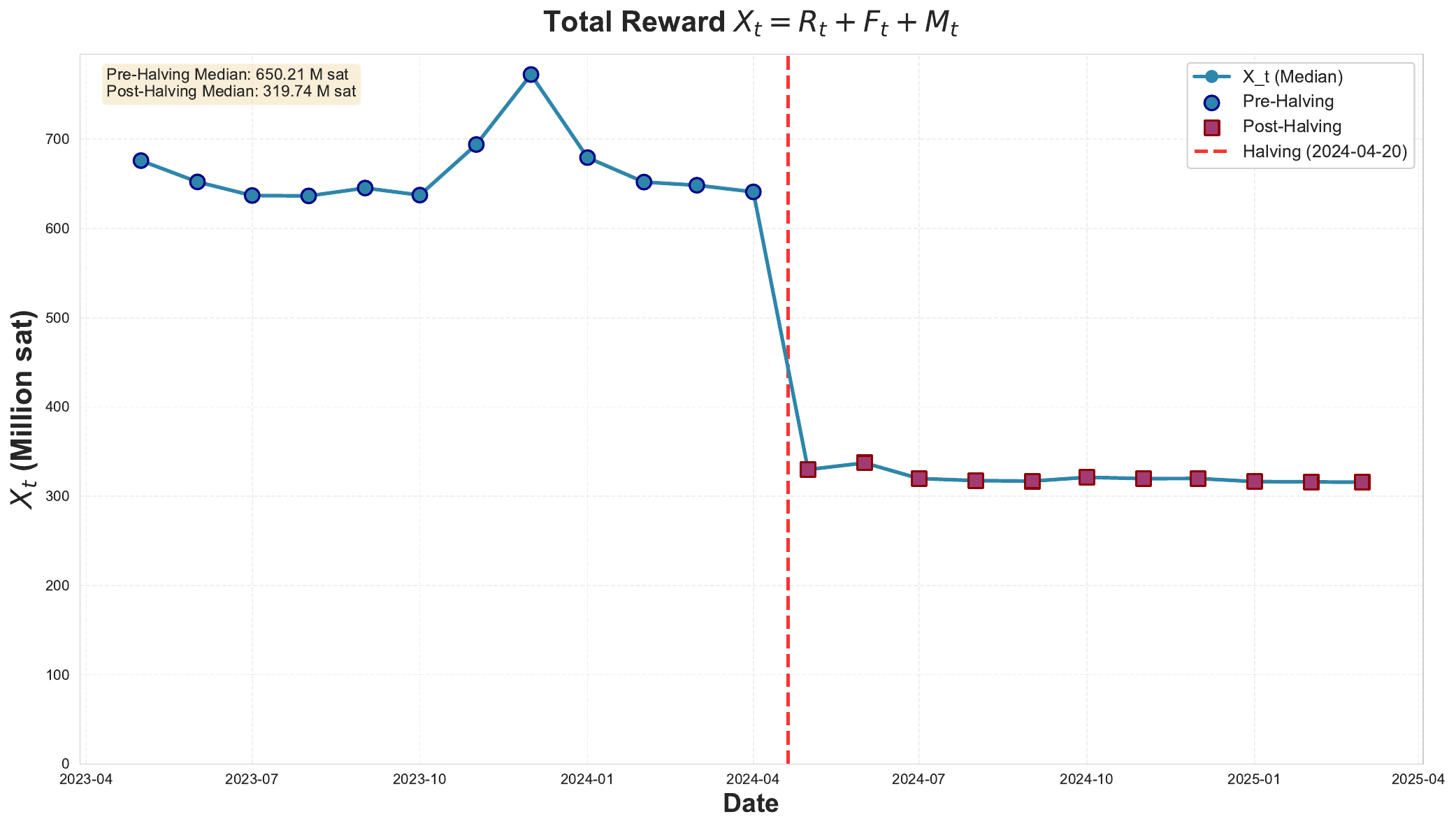}
    \caption{Miner's revenue $X_t$ changes over the 2024 halving.}
    \label{fig:Xt-over-time}

    \vspace{1cm}  

    \includegraphics[width=1\linewidth]{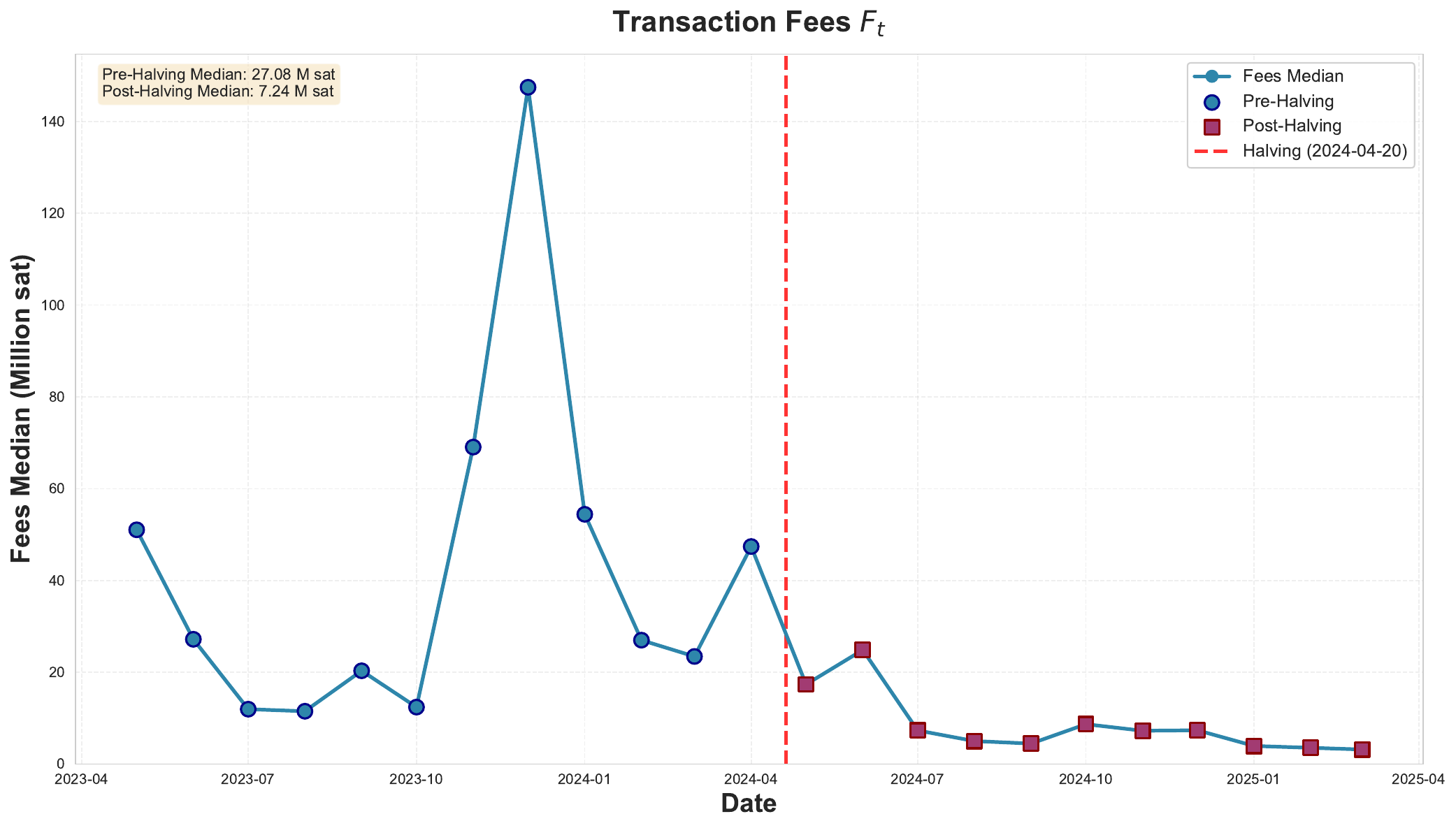}
    \caption{Transaction fees $F_t$ before and after the 2024 halving.}
    \label{fig:monthly-fees}
\end{figure}

Figure~\ref{fig:Xt-over-time} shows a substantial decline in the total observable reward $X_t$
following the 2024 halving, primarily driven by the reduction in the block reward
$R_t$ from 6.25 BTC to 3.125 BTC.
The figure indicates that the sharp decrease in $X_t$ is largely attributable to
the halving of $R_t$, while the transaction fee component $F_t$ exhibits only
moderate variation over time.

Figure~\ref{fig:monthly-fees} focuses exclusively on transaction fees and shows that the median
transaction fee level declines in magnitude after the halving.
Compared to the pre-halving period, post-halving fees display lower variability,
suggesting a sustained reduction in the fee contribution to $X_t$.

\clearpage
\FloatBarrier

\subsubsection{Impact of Reduced $X_t$ on Miner Profits $\Pi_i$}
\begin{figure}[htbp]
    \centering
    \includegraphics[width=1\linewidth]{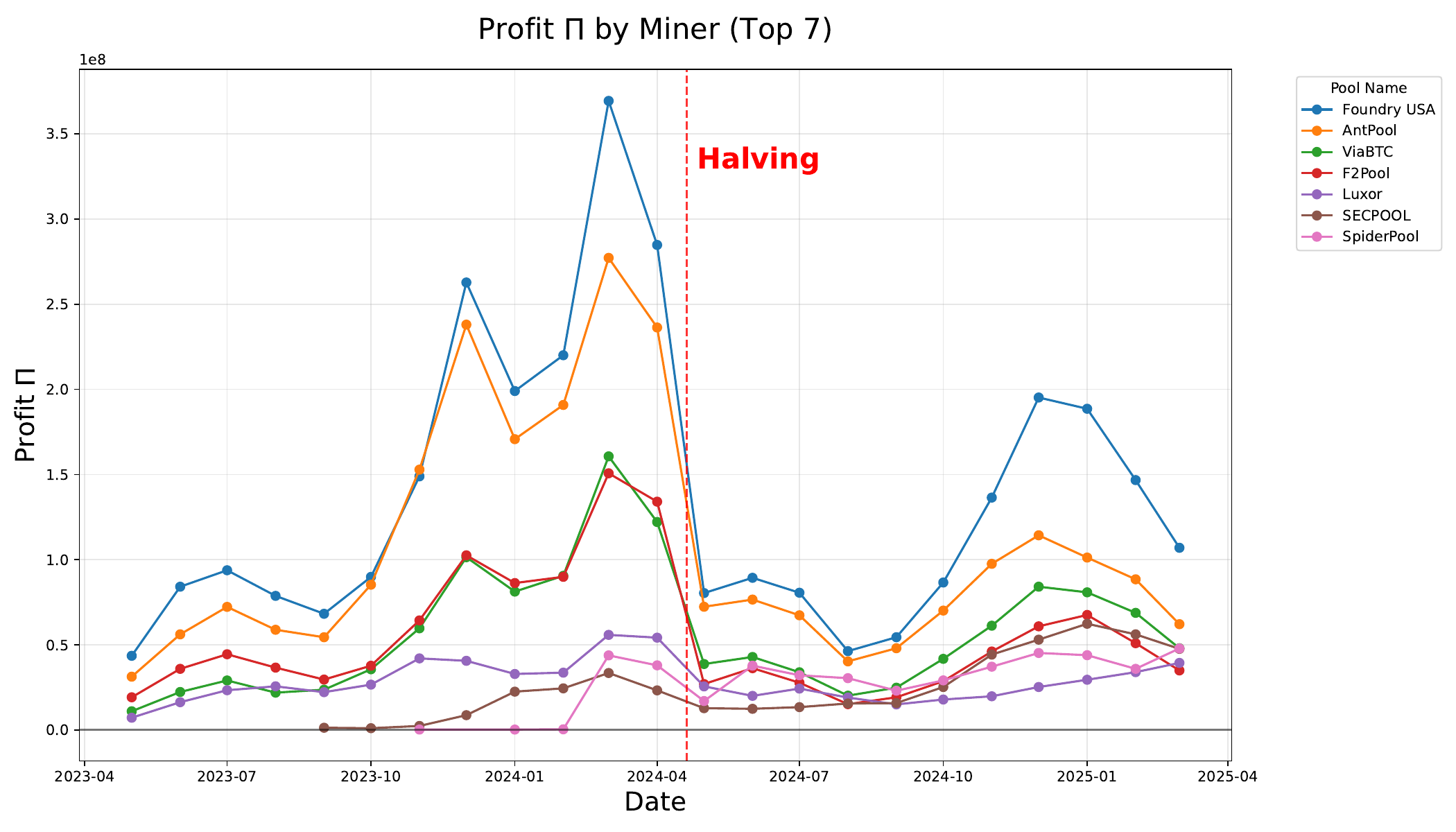}
    \caption{Monthly miner profits for major mining pools before and after the 2024 halving.}
    \label{fig:miner-profit}
\end{figure}

As shown in Figure~\ref{fig:miner-profit}, miner profit $\Pi_i$ drops sharply
following the 2024 halving, reflecting the substantial reduction in block
rewards $R_t$ and transaction fee revenue $F_t$ documented in the previous
sections.
In the post-halving period, profit recovery is concentrated among large mining
pools with substantial hashing power, while smaller pools experience persistently
depressed profit levels.
This divergence indicates that the halving-induced revenue shock is not absorbed
uniformly across miners; instead, scale plays a central role in determining both
the speed and magnitude of post-halving recovery.

These various profit outcomes imply that revenue reductions place greater
economic pressure on smaller or less efficient miners.
When such pressure persists, mining activity may gradually shift toward a smaller
number of dominant pools, as larger miners are better positioned to absorb temporary
profit shocks.
Over time, this tendency toward concentration could weaken decentralization and
pose risks to the long-run security of the network.

Looking ahead, the 2024 halving reduced the block reward from 6.25 BTC to
3.125 BTC, which still remains relatively sufficient to cover mining costs
for many participants.
However, in a future fee-only regime where $R_t = 0$, these pressures would intensify
unless they are offset by substantial increases in transaction fees or Bitcoin
price.
Moreover, in the presence of additional private gains $G_t$ from deviation-consistent
strategies, the deviation condition $G_t \ge \phi(w) \cdot X_t$ becomes
easier to satisfy as $X_t$ declines, particularly for miners with smaller hash rate
shares and higher marginal costs. In what follows, we examine whether this reduction in miner profits is associated with increased deviation-consistent behavior, even while block rewards remain positive.
\addtocontents{lof}{\protect\addvspace{10pt}} 
\FloatBarrier

\subsubsection{Shrinking On-Chain Demand after the Halving}
\begin{figure}[t]
    \centering
    \begin{subfigure}[t]{1\linewidth}
        \centering
        \includegraphics[width=\linewidth]{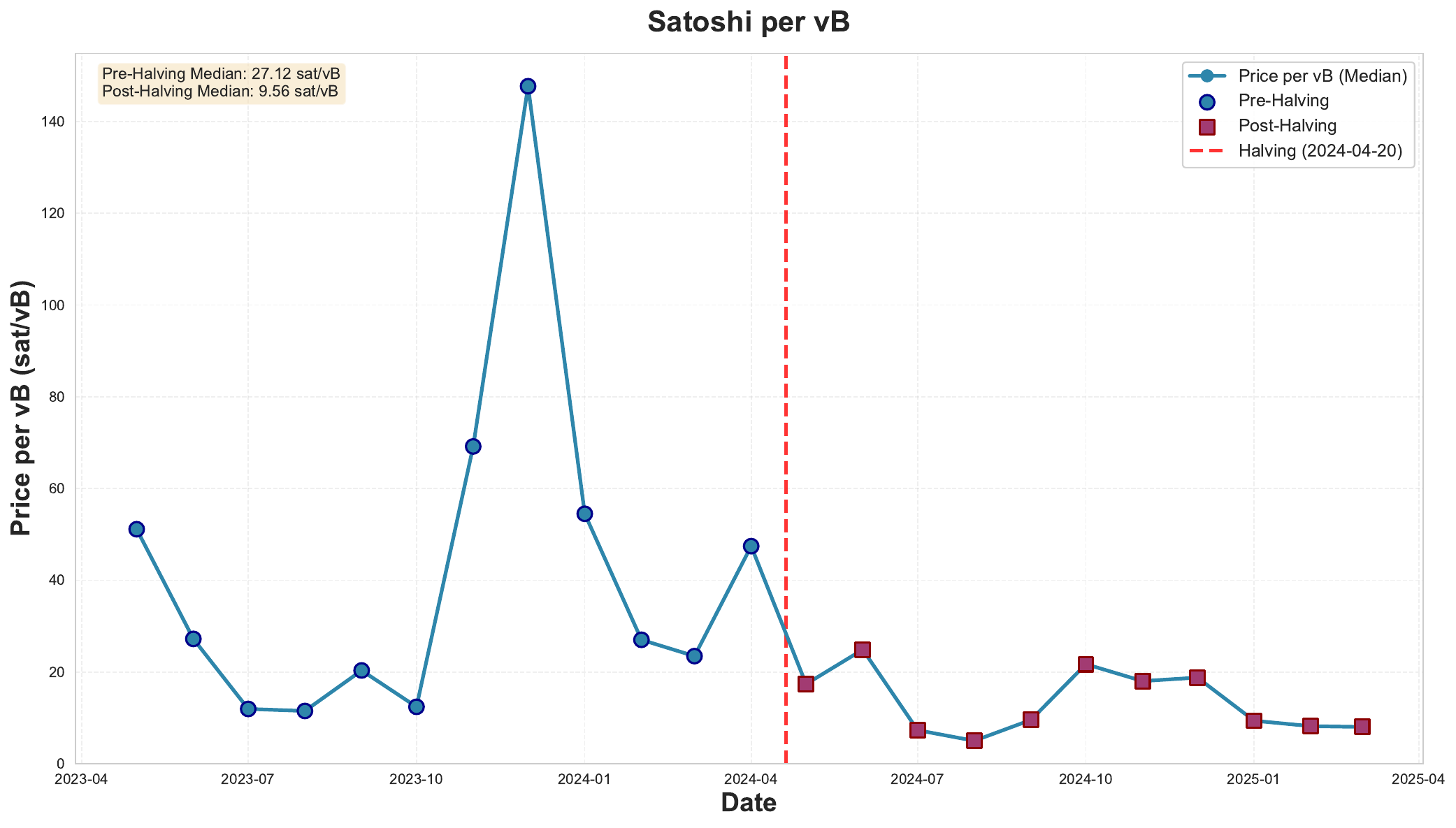}
        \caption{Transaction fee rate measured in sat/vB over time.}
        \label{fig:price-per-vb}
    \end{subfigure}
    
    \vspace{2em}
    \begin{subfigure}[t]{1\linewidth}
        \centering
        \includegraphics[width=\linewidth]{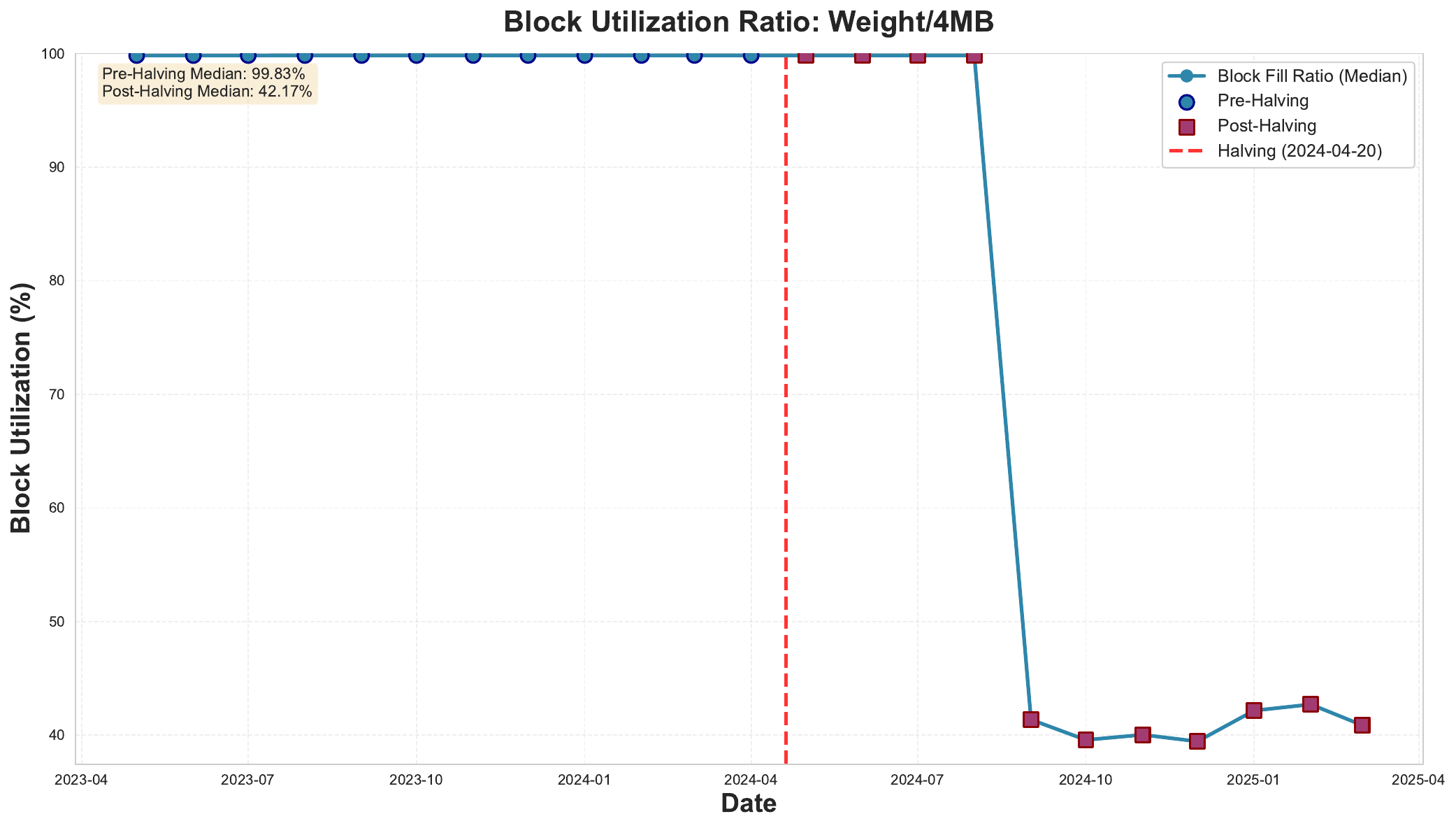}
        \caption{Monthly block utilization measured by the block fill ratio.}
        \label{fig:block-util}
    \end{subfigure}

    \caption{Post-halving decline in block space pricing and utilization.}
    
\end{figure}

Beyond the changes in miner profits discussed above, the 2024 halving is also associated with a clear contraction in on-chain demand for Bitcoin block space, reflected in both lower block space prices and reduced utilization.
As shown in Figure~\ref{fig:price-per-vb}, the median transaction fee rate (sat/vB) declines substantially in the post-halving period, indicating that block space becomes cheaper on average.

This reduction in demand is further evidenced by changes in block utilization.
Figure~\ref{fig:block-util} reports the monthly block fill ratio, which drops significantly after the halving, implying that available block space is no longer fully demanded.

Taken together, the simultaneous decline in block space prices and block utilization provides consistent evidence of shrinking on-chain demand in the post-halving regime, contributing to lower transaction fees and a reduction in the fee component of miner revenue.

\FloatBarrier

\subsubsection{Audit Scores and Deviation-Consistent Behavior}
\label{sec:deviation}
To examine whether a lower deviation threshold translates into observable changes
in miner behavior, we analyze the block audit scores provided by mempool.space.
Because miner deviation strategies are not directly observable from public blockchain data, the audit score is used as a proxy for deviation-consistent behavior. The audit score measures the ratio between the actual revenue earned by a miner and the expected revenue obtained by constructing a block using transactions from the public mempool under a standard fee-maximizing rule \cite{mempool_audit_ratio}. Lower audit scores therefore indicate departures from public-mempool-based fee maximization and are consistent with the presence of private transaction fees or other private incentives not captured by observable rewards $X_t$.

Importantly, the audit score is highly sensitive to changes in the transaction fee
market.
When transaction demand or fee-setting incentives change abruptly, the fees actually
earned by miners can differ from the fees predicted by a public-mempool-based
fee-maximizing template, even if miners do not intentionally engage in persistent
deviation strategies.
Such temporary mismatches are known to occur during periods of intensified ordinal
and inscription activity, which introduce unusual transaction patterns and alter
fee competition \cite{ordinals_analysis}.
Similar effects can also arise during episodes of heightened Bitcoin price volatility,
when transaction urgency and fee bidding behavior change rapidly \cite{flashbots2022quantifying}.

\begin{figure}[htbp]
    \centering
    \includegraphics[width=1\linewidth]{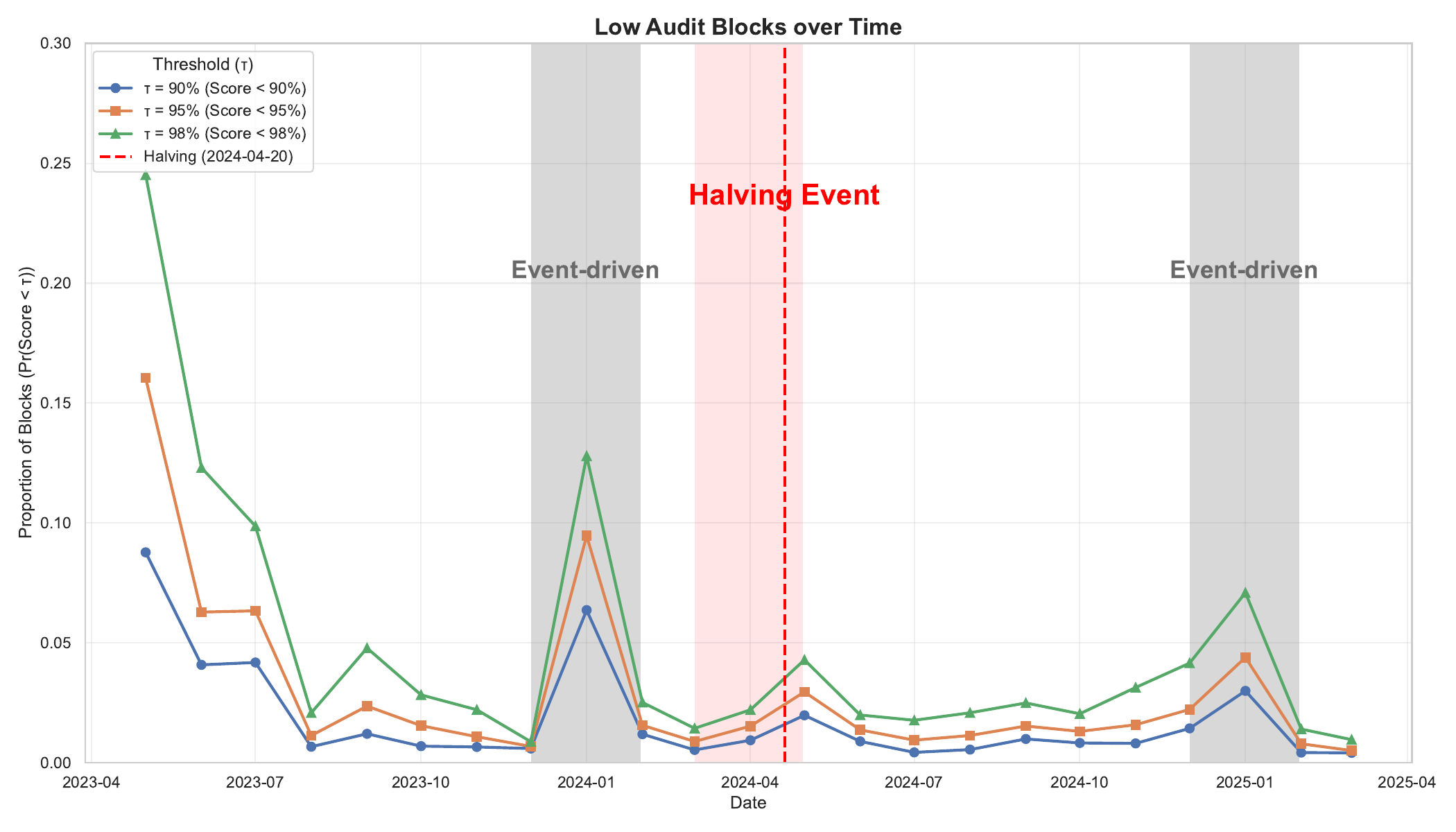}
    \caption{Time series of the proportion of blocks with low audit scores.}
    \label{fig:low-audit}
\end{figure}

Figure~\ref{fig:low-audit} presents the time series of the proportion of blocks with low audit scores. We do not observe a sustained or structural increase in low-audit score blocks following the halving event. Instead, fluctuations in audit scores are more closely associated with short-lived, event-driven disruptions in the fee market, such as changes in transaction demand induced by ordinal and inscription activity or bursts of price-driven transaction urgency.

\FloatBarrier

\subsection{Summary of Empirical Findings}
The empirical analysis shows that the 2024 halving substantially reduced observable
miner rewards by lowering the block reward $R_t$, transaction fees $F_t$, and
on-chain demand, leading to lower miner profits and a relaxation of the deviation
constraint.
Despite this relaxation, we do not find evidence of a structural or persistent
increase in deviation-consistent behavior in the post-halving period.
In particular, audit scores remain broadly stable after the halving.
If the halving had triggered widespread deviation, a sustained increase in low-audit
blocks would be expected, but such a pattern is not observed.
This stability suggests that the remaining block reward continues to play a
stabilizing role under current conditions.

Collectively, these findings indicate that the halving alone is not sufficient to
induce large-scale deviation in the short run, but deviation risks are likely to
become more pronounced as block rewards decline further toward a fee-only regime.

\section{Solutions}

\subsection{$G_t$ Threshold when $R_t = 0$}

\begin{figure}[htbp]
    \centering
    \includegraphics[width=0.75\linewidth]{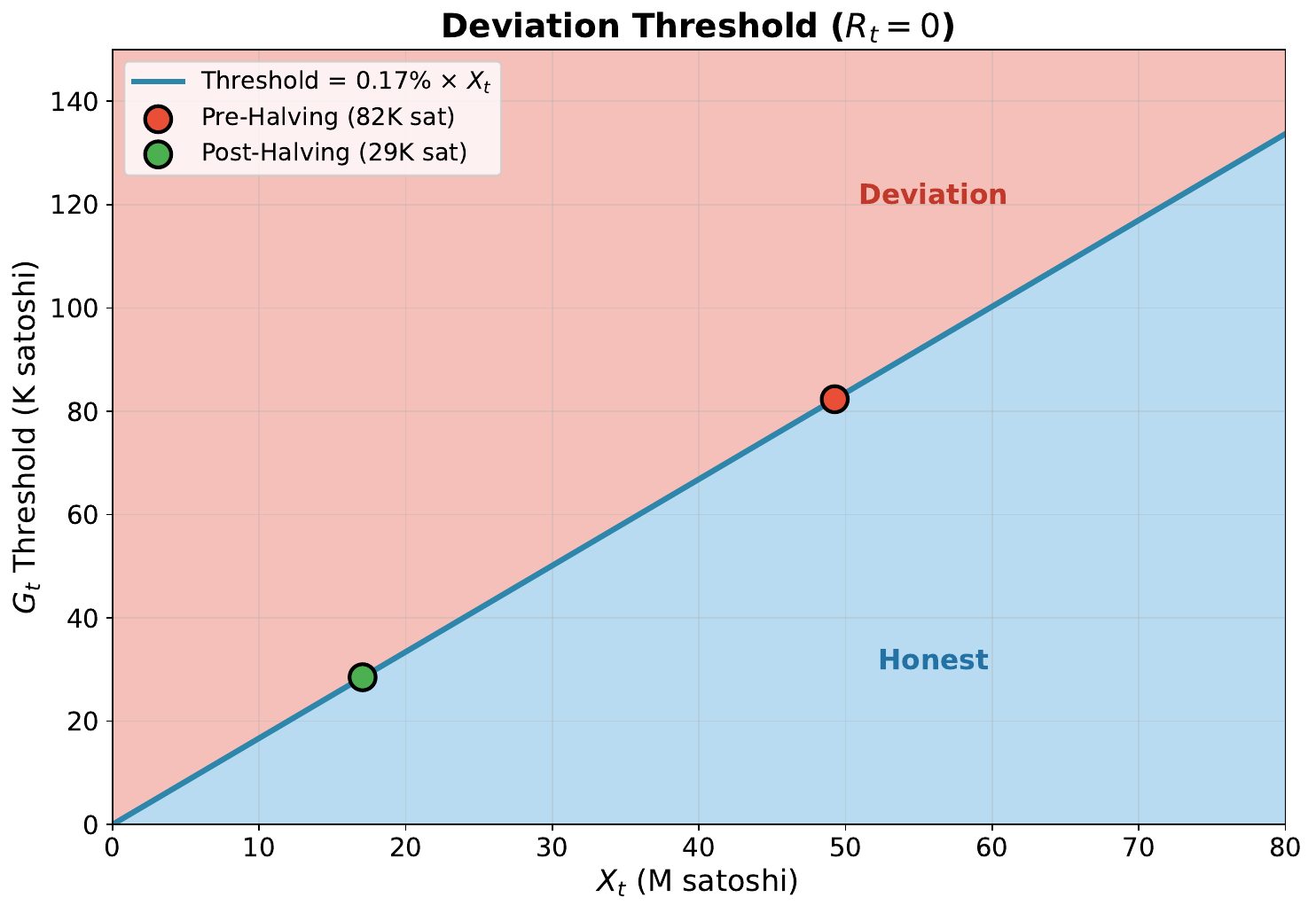}
    \caption{Additional deviation gain $G_t$ required for deviation when $R_t = 0$.}
    \label{fig:deviation-threshold-rt-zero}
\end{figure}

We consider a hypothetical fee-only regime by setting the block subsidy to zero
($R_t = 0$).
Figure~\ref{fig:deviation-threshold-rt-zero} shows the additional gain $G_t$
required for deviation as a function of the observable reward $X_t$.

The figure shows that when miner revenue comes only from transaction fees, the
deviation threshold becomes very small.
An additional gain, $G_t$, of just $0.17\%$ of $X_t$ is sufficient to make deviation
attractive across all normalized hash-rate groups in our data, since transaction
fees provide a much weaker revenue base than block rewards.
This implies that miners remain honest today mainly because block rewards still
dominate revenue, and that even small private gains may affect incentives as
block rewards decline.

\label{sec:G_t-0.17}

\subsection{Rationale Over Solution Design}

The analysis in the previous section shows that miner incentives become fragile
as the block reward declines.
This motivates a focus on the deviation threshold condition
\begin{equation}
\label{eq:deviation-threshold}
G_t \ge \phi(w) \cdot X_t,
\end{equation}
which determines whether deviation yields higher returns than honest mining, as
derived in the problem formulation chapter.

While the empirical results indicate that short-term, event-driven factors
currently explain most observed deviations, miners’ long-run behavior is
fundamentally shaped by the level and stability of the observable reward
\begin{equation}
\label{eq:observable-reward}
X_t = R_t + F_t + M_t .
\end{equation}
As the block reward $R_t$ continues to decline through future halvings and
transaction fee revenue $F_t$ becomes more volatile and demand-dependent, miner
revenue may weaken structurally, increasing the likelihood of deviation.

To prevent deviation from becoming attractive too easily, there are two broad
policy levers.
One can reduce the additional gain from deviation $G_t$, or increase the right-hand
side of the inequality by raising $\phi(w)$ or stabilizing $X_t$.
Because $G_t$ and $\phi(w)$ depend on strategy-specific and miner's specific situation, both are difficult to control directly at the protocol level.
We therefore focus on mechanisms that stabilize and support $X_t$, which is the
most tractable policy variable.

\FloatBarrier

\subsection{Base Fee and Fee Floor}

We consider a long-run regime in which the block reward converges to zero, such as after the year 2140 in Bitcoin, so that miner revenue is given by

\begin{equation}
\label{eq:fee-only-reward}
X_t = F_t + M_t.
\end{equation}
with $M_t \approx 0$.
In a fee-only regime, declining demand or reduced transaction activity can drive $F_t$ toward zero, threatening miners’ ability to cover fixed costs such as electricity and hardware depreciation.
Our objective is therefore to preserve a minimum level of miner revenue while maintaining efficient congestion control.

\subsubsection{Base Fee Design}

The base fee is a congestion-control mechanism inspired by EIP-1559 \cite{EIP1559} that adjusts transaction prices in response to realized network utilization.
Let $b_t$ denote the base fee in block $t$.
Block utilization $U_t$ is defined as the fraction of block capacity that is actually used,
\begin{equation}
\label{eq:block-utilization}
U_t = \frac{W_t}{W_{\max}},
\end{equation}
where $W_t$ denotes the realized block weight and $W_{\max}=4\,\mathrm{MB}$ is the protocol-imposed maximum under SegWit.
Let $U^*$ denote a target utilization level.

The base fee evolves according to a utilization-based feedback loop,
\begin{equation}
\label{eq:base-fee-update}
b_{t+1}
=
b_t \left(1 + \alpha \cdot \frac{U_t - U^*}{U^*}\right).
\end{equation}
where $\alpha>0$ governs the speed of adjustment.
When block utilization exceeds the target, the base fee increases to dampen demand; when utilization falls below the target, the base fee decreases to encourage transaction inclusion.
As a result, the base fee reflects congestion conditions induced by transaction demand in a disciplined manner, with $U^*$ setting the target utilization and $\alpha$ controlling the adjustment speed.

\subsubsection{Fee Floor Design}
To ensure a minimum level of transaction fees during periods of low demand, we introduce a fee floor that imposes a lower bound on effective transaction fees independent of congestion conditions.
Let $F^{\min}$ denote the fee floor.
The effective transaction fee is defined as
\begin{equation}
\label{eq:fee-floor}
F_t^{\mathrm{eff}} = \max(F_t, F^{\min}),
\end{equation}
and the effective miner reward becomes
\begin{equation}
\label{eq:effective-reward}
X_t^{\mathrm{eff}} = F_t^{\mathrm{eff}} + M_t .
\end{equation}

Unlike the base fee, which primarily mitigates fee volatility through congestion control, the fee floor directly enforces a lower bound on miner revenue.
As a result, the deviation condition becomes
\begin{equation}
\label{eq:effective-deviation-threshold}
G_t \ge \phi(w) \cdot X_t^{\mathrm{eff}}.
\end{equation}
with $X_t^{\mathrm{eff}} \ge X_t$ holding by construction.
This mechanism is particularly effective in low-demand regimes, where it prevents the deviation threshold from collapsing due to vanishing transaction fees and miner rewards.

\begin{figure}[htbp]
      \centering
      \includegraphics[width=0.7\linewidth]{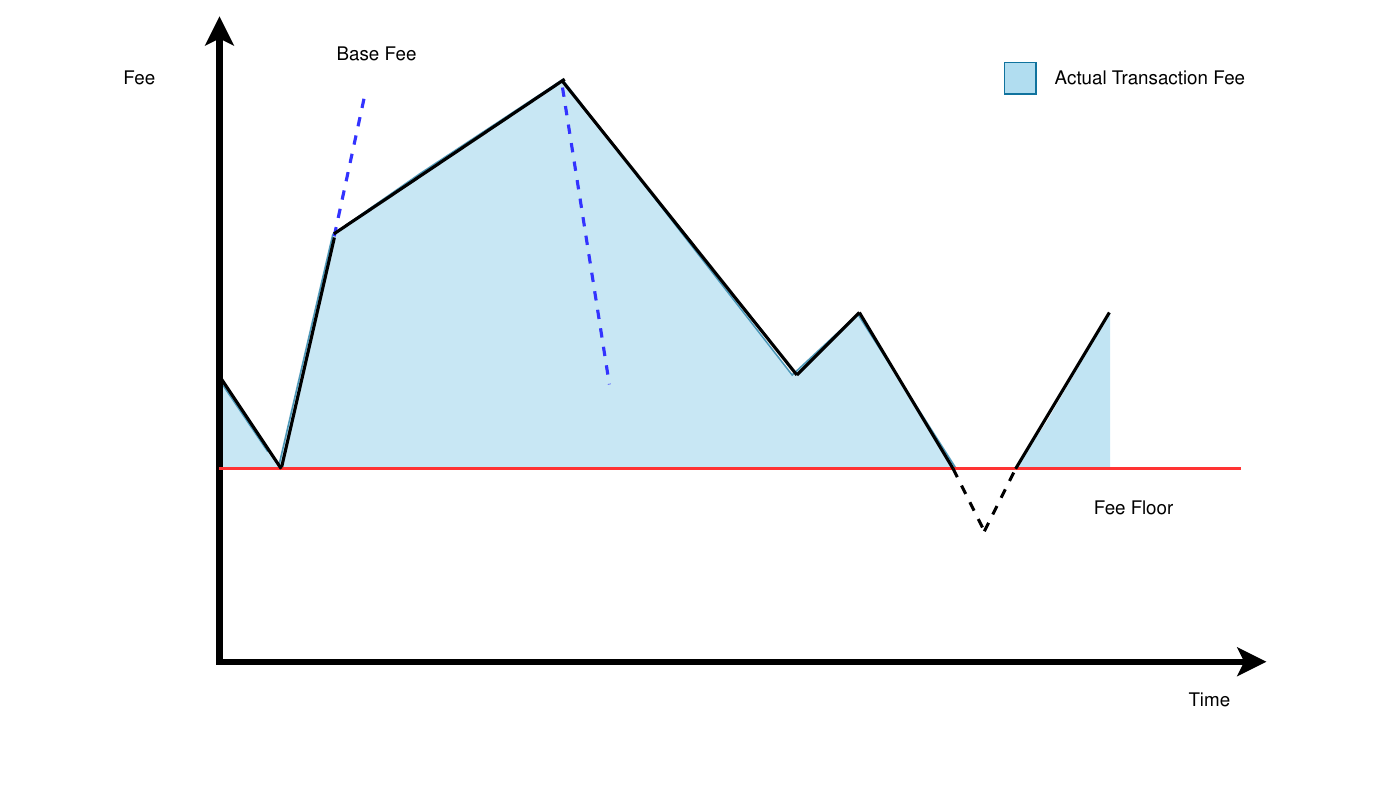}
      \caption{Base fee and fee floor example.}
      \label{fig:fee-floor}
  \end{figure}

Overall, the base fee and the fee floor serve distinct but complementary
roles. The base fee operates as a congestion-control mechanism, while the
fee floor preserves a minimum security budget by enforcing a minimum level
of miner revenue in a fee-only environment, as shown in
Figure~\ref{fig:fee-floor}.

\FloatBarrier

\subsection{Adaptive Maximum Block Size}

We propose an adaptive maximum block size policy as a supplementary mechanism for controlling persistent network congestion and miner deviation.
Rather than aiming to optimize throughput or stabilize utilization itself, the role of block size adjustment is to prevent congestion-driven incentives that increase the likelihood of miner deviation.

When block utilization remains persistently high, competition for limited
block space intensifies, making timely inclusion more expensive.
As users are forced to bid higher per-vB fees, the value of selective
inclusion and timing increases, thereby raising the potential gains from
deviating strategies:
\begin{multline}
U_t \uparrow
\;\Longrightarrow\;
\text{congestion} \uparrow
\;\Longrightarrow\;
\text{per-vB price of block space} \uparrow \\
\Longrightarrow\;
G_t \uparrow
\;\Longrightarrow\;
\text{deviation incentive} \uparrow.
\end{multline}

Conversely, when utilization remains persistently low, congestion diminishes and competition for block space weakens.
As users no longer need to bid aggressively for transaction inclusion, transaction fees decline, reducing total miner rewards:
\begin{equation}
U_t \downarrow
\;\Longrightarrow\;
\text{congestion} \downarrow
\;\Longrightarrow\;
F_t \downarrow
\;\Longrightarrow\;
X_t \downarrow
\;\Longrightarrow\;
\text{deviation incentive} \uparrow.
\end{equation}

Although the underlying mechanisms differ, both regimes relax the deviation condition by either increasing the left-hand side or reducing the right-hand side of the deviation inequality. Also, above mechanism explains why absolute bigger block size and higher network demand could not solve problem of miner's deviation. 

\subsubsection{Setting Adaptive Block Size Rule}

Instead of setting a fixed block size, we define a rule that adjusts the maximum block size in response to observed block utilization at the each epoch.
The rule updates block size conservatively while imposing explicit lower and upper bounds, $B_{\min}$ and $B_{\max}$, to prevent excessive variance in capacity.
We set the minimum block size $B_{\min}$ to 1 MB (4 million Weight Units) and the maximum $B_{\max}$ to 2 MB (8 million Weight Units).

\begin{equation}
B_{\min} \le B_t \le B_{\max}.
\end{equation}

Block size adjustments follow the utilization ratio relative to a target level.
We reuse the utilization ratio $U_t$ to compute its daily average $\bar{U}_t$ over a rolling window of 144 blocks and compare this daily average to the utilization target $U^*$ defined in the previous section. We introduce $\beta$ as a proportional adjustment factor that controls the rate of block size change.
\begin{align}
\bar{U}_t &= \frac{1}{144} \sum_{i=t-143}^{t} U_i, \\[6pt]
B_{t+1} &=
\begin{cases}
\min\{B_{\max},\,(1+\beta) B_t\}, & \text{if } \bar{U}_t > U^*, \\[6pt]
\max\{B_{\min},\,(1-\beta) B_t\}, & \text{if } \bar{U}_t < U^*, \\[6pt]
B_t, & \text{otherwise.}
\end{cases}
\end{align}

Starting from an initial baseline, the maximum block size $B_t$ evolves gradually within the interval $[B_{\min}, B_{\max}]$ using the proportional step $\beta$ at each epoch.
This conservative adjustment limits abrupt capacity changes while allowing the network to respond to persistent congestion.
In doing so, the policy avoids excessive increases in deviation gains $G_t$ caused by congestion-induced delay, and prevents fee collapse that would otherwise reduce effective miner rewards.

\subsection{Simulation Results}

We evaluate miner deviation behavior under combinations of three policy
mechanisms: Base Fee, Fee Floor, and Adaptive Block Size, in a zero block reward
setting ($R_t = 0$).
Historical Bitcoin blocks from heights 790{,}000 to 890{,}000 are used to assess
the effectiveness of these policies.
For each block, miners compare profits from honest and deviating strategies
using observed transaction fees, block size, MEV estimates, and miner cost data,
consistent with the empirical data analysis in the previous section.

Miner deviation is evaluated using the $G_t$ threshold, and the deviation rate
$\beta$ is computed as the fraction of blocks in which miners choose deviation.
Following the Bitcoin Backbone framework~\cite{bitcoin_backbone,DBLP:conf/crypto/GarayKL17}, network
stability requires the deviation rate $\beta$ to remain below 50\%.
Detailed simulation assumptions, parameter grids, and robustness checks are reported in the Appendix.

\begin{figure}[htbp]
    \centering
    \includegraphics[width=1\linewidth]{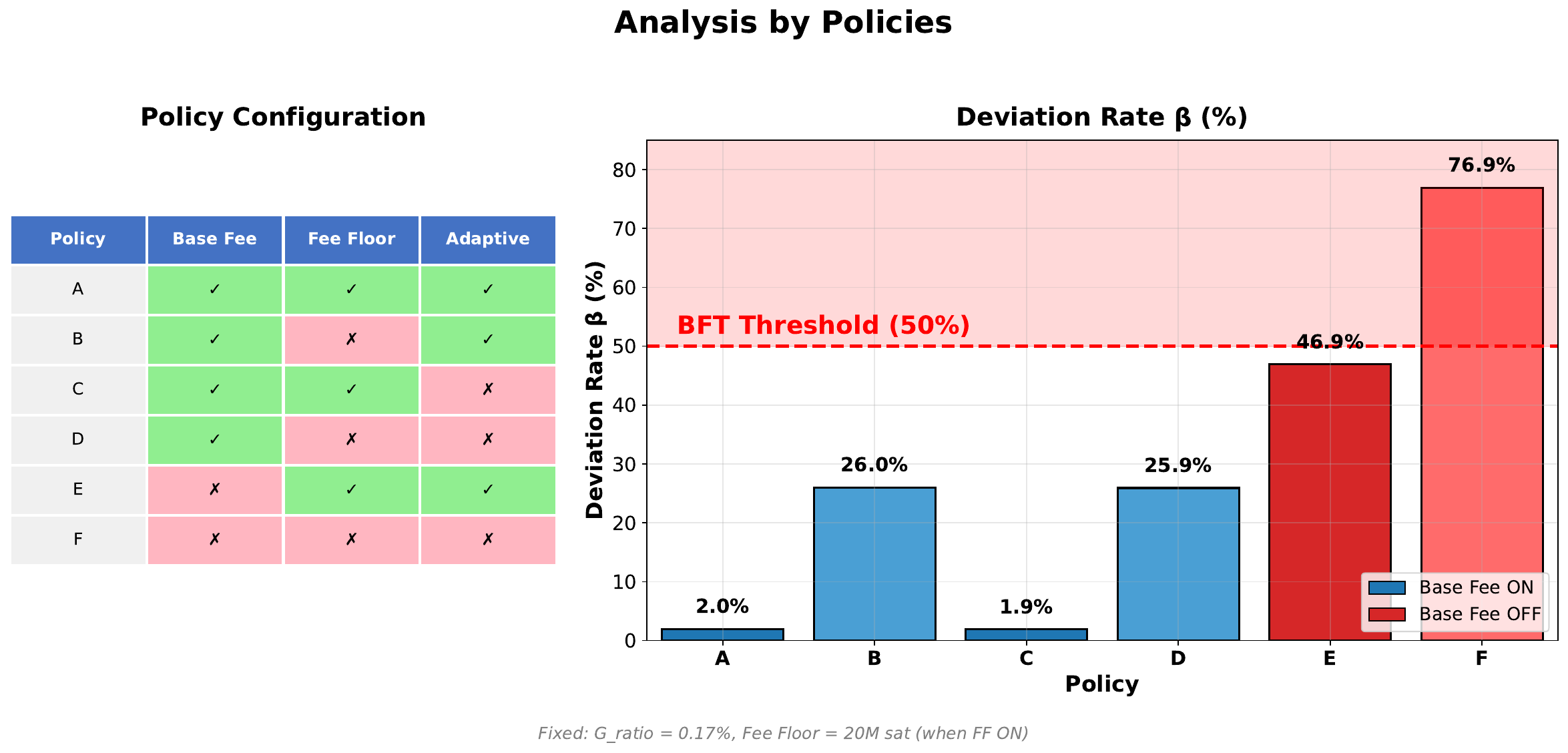}
    \caption{Deviation rates under different policy combinations.}
    \label{fig:solution_table}
\end{figure}

Figure~\ref{fig:solution_table} summarizes miner deviation outcomes across all
policy configurations under a zero block reward scenario, with the additional
gain fixed at $0.17\%$ of $X_t$, where deviation becomes relatively easy, as
discussed in Section~\ref{sec:G_t-0.17}.
In policy configuration~F, where no policy is applied, deviation becomes
widespread even for this small additional gain, with the deviation rate exceeding
70\%.
This indicates that a fee-only regime without stabilization mechanisms is highly
fragile to deviation.

Introducing policy mechanisms substantially reduces deviation.
Every configuration that retains the base fee, namely A, B, C, and D,
remains within the BFT threshold, and the strongest performance is
observed in A and C, where the base fee is paired with the fee floor;
extending this pairing with adaptive block size, as in A, produces only a marginal difference from C.
Policy E, which omits the base fee, also stays within the threshold,
but its deviation rate approaches the limit, confirming that the base fee
is the dominant factor in preventing deviation.
Policy F, with no policy applied, exceeds the threshold and
illustrates the fragility of a fee-only regime without stabilization.
These results show that stabilizing observable miner rewards is essential
for maintaining security in a post-block-reward environment, with the base
fee providing the foundational guarantee and the fee floor reinforcing it
most effectively.

\section{Further Discussion}

In the problem formulation section, the withholding penalty factor
$\phi(w)$ was defined as
\begin{align}
\phi(w)
\triangleq
\frac{p_i^{\mathrm{hon}} - p_i^{\mathrm{dev}}}{p_i^{\mathrm{dev}}},
\label{eq:ratio-def}
\end{align}
which measures the relative difference in mining success probability
between honest and deviating strategies.
Under standard assumptions on block discovery and propagation,
the mining success probability can be written as
\begin{align}
p_i = h_i \cdot (1 - \rho),
\label{eq:success-prob}
\end{align}
where $h_i$ denotes miner $i$'s hash rate share and $\rho$ denotes the
orphan probability, reflecting the likelihood that a mined block fails
to be successfully propagated.

Substituting this expression into the definition of $\phi(w)$
yields
\begin{align}
\phi(w)
&= \frac{h_i (1 - \rho^{\mathrm{hon}}) - h_i (1 - \rho^{\mathrm{dev}})}
        {h_i (1 - \rho^{\mathrm{dev}})} \notag \\
&= \frac{\rho^{\mathrm{dev}} - \rho^{\mathrm{hon}}}
        {1 - \rho^{\mathrm{dev}}},
\label{eq:ratio-network}
\end{align}
which shows that $\phi(w)$ is independent of the individual miner's 
hash rate and is instead determined by differences in orphan probabilities 
across mining strategies. Since the orphan rate gap between honest and deviating mining strategies arises from the withholding time $w$, we can simplify the equation as follows.

\begin{align}
\phi(w)
&= \frac{\rho^{dev}-\rho^{hon}}{1-\rho^{dev}} \\
&= \frac{e^{-\lambda \delta^{hon}}\left(1 - e^{-\lambda w}\right)}
       {e^{-\lambda(\delta^{hon}+w)}} \\
&= \frac{1 - e^{-\lambda w}}{e^{-\lambda w}} \\
&= e^{\lambda w} - 1.
\end{align}

As a result, a systematic analysis of how orphan rates and the withholding 
time $w$ interact with the deviation threshold $G_t$ is left for future work.

\section{Conclusion}
This work began by examining miner profitability and its role in preventing
strategic deviation in the long run.
Bitcoin is the first decentralized currency that enables permissionless
participation and financial autonomy without reliance on centralized authority.
As such, the stability of its incentive structure is central to the success of
this ongoing social and economic experiment.

To date, the Bitcoin network has not experienced a hard fork driven by
self-interested behavior of nodes or large-scale miner deviation.
However, as block rewards decline and eventually vanish, miner incentives may
weaken, potentially increasing the vulnerability of the network even before a
pure fee-only regime is reached.
A loss of confidence in network security could further amplify this risk.

Our analysis shows that protocol-level mechanisms such as the base fee, fee
floor, and adaptive block size can play an important role in stabilizing miner
incentives by supporting observable rewards.
While these mechanisms are not presented as definitive solutions, they highlight
key design directions and raise important questions for stakeholders concerned
with the long-term security and sustainability of Bitcoin.

\begin{credits}
  \subsubsection{\ackname}
  The author thanks his thesis advisor, Prof.~Juan A.~Garay
  (Department of Computer Science and Engineering, Texas A\&M
  University), for his invaluable guidance and feedback
  throughout this work, including the comments received during
  the thesis defense.

  \subsubsection{\discintname}
  The author has no competing interests to declare that are
  relevant to the content of this article.
\end{credits}

\renewcommand{\refname}{References}
\providecommand{\bibname}{}\renewcommand{\bibname}{References}
\bibliographystyle{splncs04}
\bibliography{refs}

\appendix
\input{appendix}

\end{document}

%% file: appendix.tex
\newpage
\section{Additional Proofs on Assumptions and the Solutions}
\subsection{$G_t$ Threshold and $\phi(w)$}

In this appendix, we formally show that the successful mining probability gap 
$(p_i^{hon} - p_i^{dev})$ appearing in the $G_t$ threshold condition 
and the orphan rate gap $(\rho^{dev} - \rho^{hon})$ in 
$\phi(w)$ are non-negative. We assume:
\begin{itemize}
\item Withholding time $w \ge 0$,
\item Block arrival rate $\lambda > 0$ (Bitcoin targets one block per 600 seconds),
\item Hash share $h_i > 0$.
\end{itemize}

\subsubsection{Orphan Rate}

The orphan probability under propagation delay $\delta$ is
\begin{align}
\rho(\delta) = 1 - \exp(-\lambda \delta).
\end{align}

\subsubsection{Successful Mining Probability}
The successful mining probability of miner $i$ is
\begin{align}
p_i(\delta) 
&= h_i (1 - \rho(\delta)) \\
&= h_i \exp(-\lambda \delta).
\end{align}

\subsubsection{Propagation Delay under Deviation}
For a deviating miner with withholding time $w \ge 0$,
\begin{align}
\delta^{dev} = \delta^{hon} + w.
\end{align}

\subsubsection{Gap in Successful Mining Probability}
\begin{align}
p_i^{hon} 
&= h_i \exp(-\lambda \delta^{hon}), \\
p_i^{dev} 
&= h_i \exp(-\lambda \delta^{dev}) \\
&= h_i \exp\!\bigl(-\lambda(\delta^{hon}+w)\bigr) \\
&= p_i^{hon}\exp(-\lambda w).
\end{align}

Therefore,
\begin{align}
p_i^{hon} - p_i^{dev}
&= p_i^{hon}\bigl(1 - \exp(-\lambda w)\bigr).
\end{align}

Since $w \ge 0$ and $\lambda > 0$,

\begin{align}
0 < \exp(-\lambda w) \le 1
\quad \Rightarrow \quad
1 - \exp(-\lambda w) \ge 0.
\end{align}

Hence,

\begin{align}
p_i^{hon} - p_i^{dev} \ge 0.
\end{align}

The inequality is strict when $w > 0$. 
When $w = 0$, the deviating strategy is identical to honest mining, 
as no additional withholding delay is introduced. 
In this case, the deviation condition reduces to a trivial comparison, 
and the $G_t$ threshold no longer carries substantive meaning. As a result, the gap in successful mining probability in $G_t$ threshold is non-negative.

\subsubsection{Gap in Orphan Rate}
\begin{align}
\rho^{hon} &= 1 - \exp(-\lambda \delta^{hon}), \\
\rho^{dev} &= 1 - \exp(-\lambda \delta^{dev}) \\
           &= 1 - \exp\!\bigl(-\lambda(\delta^{hon}+w)\bigr).
\end{align}

Taking the difference,

\begin{align}
\rho^{dev} - \rho^{hon}
&= \exp(-\lambda \delta^{hon})
   - \exp\!\bigl(-\lambda(\delta^{hon}+w)\bigr) \\
&= \exp(-\lambda \delta^{hon})
   \left(1 - \exp(-\lambda w)\right).
\end{align}

Since $\exp(-\lambda \delta^{hon}) > 0$ and 
$1 - \exp(-\lambda w) \ge 0$,

\begin{align}
\rho^{dev} - \rho^{hon} \ge 0.
\end{align}

Thus, withholding time $w$ weakly increases the orphan probability while weakly decreasing 
the successful mining probability. Consequently, the orphan rate gap 
underlying $\phi(w)$, as discussed in the Further Discussion chapter, 
is non-negative.

\section{Simulation Parameters and Data Sources}
\subsection{Reproducibility}

All code, configuration files, and data-processing scripts needed to reproduce
the experiments and figures in this thesis are publicly available in the
project repository:
\begin{center}
\url{https://github.com/xodn348/BTC_EXP}
\end{center}
The repository includes (i) end-to-end data processing pipelines, (ii)
simulation code and default configuration files, and (iii) scripts to
regenerate the main experimental outputs.

\subsection{Network and Protocol Parameters}

\begin{table}[h]
\centering
\caption{Network and Protocol Parameters}
\label{tab:network_params}
\begin{tabular}{llll}
\toprule
\textbf{Parameter} & \textbf{Symbol} & \textbf{Value} & \textbf{Source/Rationale} \\
\midrule
Block arrival rate & $\lambda$ & $1/600~\mathrm{s}^{-1}$ & Bitcoin protocol (10-minute target) \\
Discount factor & $\gamma$ & 0.99993 & Standard in repeated-game models \\
Base network delay & $\delta_0$ & 742 ms & KIT invstat.gpd (50th percentile) \\
Delay per MB & $\kappa$ & 26.40 ms/MB & Linear regression on propagation data \\
Withholding delay & $w$ & 1.0 s & Eyal \& Sirer (2014) \\
\bottomrule
\end{tabular}
\end{table}

\newpage
\subsection{Policy Mechanism Parameters}

\begin{table}[h]
\centering
\caption{Policy Mechanism Parameters}
\label{tab:policy_params}
\begin{tabular}{llll}
\toprule
\textbf{Parameter} & \textbf{Symbol} & \textbf{Value} & \textbf{Description} \\
\midrule
Base fee adjustment speed & $\alpha$ & 0.125 & EIP-1559--inspired update rate \\
Initial base fee & $b_0$ & 20 sat/vB & Median observed fee rate \\
Target utilization & $U^*$ & 0.80 & Target block utilization \\
Block size adjustment step & $\beta$ & 0.10 & Per-epoch adjustment rate \\
Minimum block size & $B_{\min}$ & 1 MB & Lower bound for adaptive sizing \\
Maximum block size & $B_{\max}$ & 2 MB & Upper bound for adaptive sizing \\
\bottomrule
\end{tabular}
\end{table}

\FloatBarrier
\subsection{MEV Modeling Parameters}
MEV is generated using a zero-inflated lognormal distribution, where a fraction
of blocks contain zero MEV and non-zero realizations are drawn from a lognormal
distribution parameterized by $(\mu, \sigma)$.
The parameter values are adapted from empirical MEV studies in Ethereum and
the BlockScholes modeling framework, and rescaled to reflect Bitcoin's
lower MEV prevalence~\cite{blockscholes2024mev}.
\begin{table}[h]
\centering
\caption{MEV Distribution Parameters}
\label{tab:mev_params}
\begin{tabular}{lll}
\toprule
\textbf{Parameter} & \textbf{Value} & \textbf{Source/Rationale} \\
\midrule
Zero-inflation rate & 0.80 & Adjusted from Ethereum MEV measurements \\
Lognormal mean ($\mu$) & 14.9 & Calibrated to match target MEV magnitude \\
Lognormal std.\ dev.\ ($\sigma$) & 1.8 & Captures long-tail behavior \\
Maximum MEV & $0.1 R_t$ & Conservative upper bound \\
\bottomrule
\end{tabular}
\end{table}

\subsection{Data Sources and Processing}
Simulations are conducted using the top 13 mining pools by average hash
rate share, covering approximately 99.5\% of the total network hashing power.
Pool-level mining costs are computed using CBECI's annualised electricity
consumption estimate (GUESS), converted to a daily network-wide cost via
an assumed electricity price, and then allocated to each pool based on
the number of blocks it actually mined on that day.

\begin{table}[h]
\centering
\caption{Simulation Data Sources and Processing}
\label{tab:data_sources_detailed}
\begin{tabular}{lll}
\toprule
\textbf{Category} & \textbf{Data} & \textbf{Source} \\
\midrule
On-chain blocks
& Block height, fees, vbytes, tx count, pool ID
& Blockchain.com API \\

Price data
& BTC--USD daily price
& Yahoo Finance \\

Mining costs
& Annualised electricity consumption (TWh/year)
& Cambridge CBECI \\

Pool shares
& Daily pool hashrate shares
& Blockchain.com API \\

MEV (synthetic)
& MEV samples (sat, USD)
& Parameter-based model \\

Propagation delay
& Block propagation statistics
& KIT invstat.gpd \\

\bottomrule
\end{tabular}
\end{table}
